\documentclass{iopart}
\usepackage{graphicx}
\usepackage{latexsym}

\newcommand{\be}{\begin{equation}}
\newcommand{\ee}{\end{equation}}
\newcommand{\bea}{\begin{eqnarray}}
\newcommand{\eea}{\end{eqnarray}}

\begin{document}
\title{Abundance of unknots in various models of polymer loops}

% this is the Phys Rev E header
%\author{N.T. Moore}
%\email{nmoore@winona.edu}
%\altaffiliation[Now at ]{Department of Physics, Winona State University, Winona, MN 55987, USA}
%\author{A.Y. Grosberg}
%\email{grosberg@physics.umn.edu}
%\affiliation{Department of Physics, University of Minnesota, Minneapolis, MN 55455, USA}
%\date{\today}

% This the the IOP JPhysA header
\author{N.T. Moore$^1$ and A.Y. Grosberg$^2$}
\address{$^1$ Physics Department, Winona State University, Winona, MN 55987, USA.}
\address{$^2$ Department of Physics, University of Minnesota, Minneapolis, MN 55455, USA}
\ead{$^1$ ntmoore@gmail.com}
\ead{$^2$ grosberg@physics.umn.edu}
\date{\today}

\begin{abstract}
A veritable zoo of different knots is seen in the ensemble of
looped polymer chains, whether created computationally or observed
in vitro. At short loop lengths, the spectrum of knots is
dominated by the trivial knot (unknot).  The fractional abundance
of this topological state in the ensemble of all conformations of
the loop of $N$ segments follows a decaying exponential form, $
\sim \exp \left( -N/N_0 \right)$, where $N_0$ marks the crossover
from a mostly unknotted (ie topologically simple) to a mostly
knotted (ie topologically complex) ensemble. In the present work
we use computational simulation to look closer into the variation
of $N_0$ for a variety of polymer models.  Among models examined,
$N_0$ is smallest (about $240$) for the model with all segments of
the same length, it is somewhat larger ($305$) for Gaussian
distributed segments, and can be very large (up to many thousands)
when the segment length distribution has a fat power law tail.
\end{abstract}

% guessing at PACS code,
%   57M25 Knots and links in $S^3$ {For higher dimensions, see 57Q45}
%   82D60 Polymers
%\pacs{57M25,82D60}
%\submitto{\JPA}
\maketitle

\section{Introduction: Formulation of the Problem}
Of interest to anglers seeking to fill their creels and children
seeking to fasten their shoes, a wide audience has found knots compelling from
time immemorial.  In the scientific community, knots have been
featured in initial formulations of the nature of atoms,
\cite{knotted_vortices} (see a popular historical account in
\cite{historical_paper}), the formulation of certain path
integrals, \cite{Jones_polynomial_applications}, and also in
quantitative biology, where knots have been observed in,
\cite{JBiolChem_1985,probability_DNA_knotting}, and tied into, DNA,
\cite{tie_knot_into_DNA-Japan,tie_knot_into_DNA-Quake}, where the
space of knots is biologically created and manipulated \cite{space_of_knots}. 
Knots also have been observed occasionally in proteins,
\cite{Mansfield_correspondance,Taylor_knots_in_proteins,
trefoil_in_protein,trefoil_in_protein_2,Rhonald_proteins,Kardar_protein_knot}.

Historically, the classification of knots and study of knot invariants were the
first subjects of knot theory \cite{historical_paper}, 
%and this mathematical interest persists to the present day, 
and this remains in the center of attention among knot theorists of mathematical orientation
\cite{Jones_polynomial_applications}. 
Another fundamental aspect of knot theory is that of knot entropy.
Physically, this group of problems comes to the fore in the context of polymers 
and biophysics. Mathematically, this issue belongs to both topology and
probability theory and seems to remain underappreciated in the
mathematics and mathematical physics community.  
Even the simplest question in this area is poorly
understood: what is the probability that a randomly closed loop in
$3D$ will be topologically equivalent to plane circle? In other
words, using professional parlance of the field, what is the
probability that random loop is a trivial knot (unknot), $0_1$?  
There are, of course, many more questions along the same
lines, e.g., what are probabilities of other more complex knots?
what is the entropic response of a topologically constrained loop
to various perturbations, etc.

Most of what we know about these ``probabilistic topology''
questions is learned from computer simulations.  In particular, it
has been observed by many authors over the last 3 decades
\cite{maximfirst,koniaris_muthu_N0,Deguchi_universality_1997,Lattice_knot_probability,Vologodskii_hedgehog}
that the trivial knot probability depends on the length of the
loop, decaying exponentially with the number of segments in
the loop, $N$:
\begin{equation}
w_{\rm triv} = A \exp \left( -N/N_0 \right) \ . \label{eq:triv_P}
\end{equation}
For some lattice models this exponential law, in the $N \to \infty$ 
asymptotics, was also mathematically proven 
\cite{proof_of_exponential_abundance_1,proof_of_exponential_abundance_2}.
It was also noticed \cite{Deguchi_universality_1997} that the same
exponential law, with the same decay parameter $N_0$, also describes
the large $N$ asymptotical tail of the abundance of any other particular knot - although for
complex knots exponential decay starts only at sufficiently large $N$ 
(as soon as the given knot can be identified as an underknot \cite{Nathan_Chapter_19}).

An alternative view of formula (\ref{eq:triv_P}), useful in the context of thermodynamics,
implies that the removal of all knots from the loop is associated with
thermodynamically additive (linear in $N$) entropy loss of $1/N_0$
per segment; in other words, at the temperature $T$, untying all
knots would require mechanical work of at least $k_BT/N_0$ per
segment.

Another manifestation of the importance of the $N_0$ parameter was
found in the recent series of works
\cite{AG_pred,Nathan_PNAS,Nathan_Chapter_19,Nathan_PRE}.  These
works belong to the direction
\cite{desCloizeaux_conj,Quake1,Deutsch,Deguchi_2003,swiss_PNAS}
addressing the spatial statistics of polymer loops restricted to
remain in a certain topological knot state.  It turns out that
even for loops with no excluded volume and thus are not
self-avoiding, $N_0$ marks the crossover scale between mostly
Gaussian ($N<N_0$) and significantly non-Gaussian ($N>N_0$)
statistics.  Indeed, at $N<N_0$, locking the loop in the state of
an unknot excludes only a small domain of the conformational space
which produces only marginal (albeit non-trivial
\cite{Nathan_PRE}) corrections to Gaussian statistics - for
instance, mean-squared gyration radius of the loop is nearly
linear in $N$.  By contrast, at $N>N_0$, the topological constraints
are of paramount importance, making the loop statistics very much
non-Gaussian, and consistent with effective self-avoidance
\cite{Nathan_PNAS,Nathan_Chapter_19,desCloizeaux_conj,swiss_PNAS}.

Thus, it seems likely that the parameter $N_0$ might hold the key
to the entire problem of knot entropy.  We therefore decided to
look at this parameter more closely in this paper.

Present understanding of the values of $N_0$ is quite modest. 
First, the constant's value was invariably found to be quite large, 
around $300$ for all examined models of ``thin'' loops with 
no excluded volume, or no self-avoidance
\cite{maximfirst,koniaris_muthu_N0,Deguchi_universality_1997,Nathan_PNAS,Nathan_Chapter_19}.
Second, it is known that knots are dramatically suppressed for
 ``thick'' self-avoiding polymers, which means that $N_0$ rapidly increases with
the radius of self-avoidance \cite{Deguchi_universality_1997,Vologodskii_hedgehog}. 
The latter issue is also closely connected to the probabilities of
knots in lattice models, where the non-zero effective
self-avoidance parameter is automatically set by the lattice
geometry.  In the present paper, we will only consider the
arguably more fundamental case of loops with no self-avoidance.

The starting point of our analysis is the observation that $N_0$
appears to be noticeably different for two standard polymer models
for which common sense suggests that they should be equivalent.
Both models can be called freely-jointed in the sense that they
consist of $N$ rigid segments with free rotation in the joints.
However, in one model all segment vectors are of the same length,
while in the other model segment vectors are taken from a Gaussian
distribution. The motivation to consider the Gaussian distributed
step vectors comes from the idea of decimation, or
renormalization: we can start from the loop with $Ng$ segments of
equal length and then group them into $N \gg 1$ blobs of $g \gg 1$
bare segments, each blob having nearly Gaussian distributed
end-to-end vector.  With respect to the knot abundance, the fixed
length model was examined in \cite{koniaris_muthu_N0} and the
Gaussian model in \cite{Deguchi_universality_1997}.  It was
noticed that $N_0$ for the Gaussian distributed steps was larger
than for identical steps, assuming no self-exclusion in both
cases.  No attention was paid to this observation, possibly
because there was no confidence that the observed difference is
real, in the context of the numerical error bars in the pertinent measurements.

Recently, \cite{Nathan_PNAS,Nathan_PRE}, more detailed data became
available which suggest that indeed $N_0$ is different for the two
models, with fixed or Gaussian distributed steplength. A similar
result was independently obtained by Vologodskii
\cite{Pittsburgh_talk}.  Latter in this article, we present even
better quality data supporting the same observation that $N_0$ is
different for these two models.  This is a rather disturbing
observation. Indeed, the idea of universality in polymer physics
\cite{DeGennesBook} suggests that there should not be any
difference between these two models as far as any macroscopic
quantity is concerned.  For instance, not only is the mean squared
gyration radius the same for both models, but even the
distribution of the gyration radii are the same, except far into
the tails.  In general, the difference between polymer models of
this type becomes significant only in the strong stretching regime
\cite{MarkoSiggia} or at high density \cite{AG_Red}.  Even if one
takes into account the idea that knots, when present, are most
likely localized along the chain 
\cite{Rhonald_Chapter_20,localization_1,localization_2,localization_3,localization_4}, 
it is unclear how this fact can manifest itself for the loop that has no knots.

Thinking generally about the loop models with fixed or Gaussian
steplength, our reaction to this discrepancy is to realize that
the major difference between the two freely-jointed loop models
is that the Gaussian model may have a few unusually long segments,
suppressing the ability of other shorter segments to wind around,
and thus decreasing the possibility for knots to occur, and accordingly,
increasing $N_0$.
Thus, the ability to take long strides might account for the
comparative slowness of the ensemble of loops with Gaussian
steplengths to diversify their knot spectrum with increasing $N$.
The main goal of the present work is to investigate this
conjecture.

The plan of the work is as follows.  After a brief description of
our computational algorithms used to generate closed loops and to
identify their topologies (section \ref{sec:Methods}), we present
computational results (section \ref{sec:Results}) on knot
abundance for a variety of models differing in the width of their
steplength distribution. In addition to the already mentioned
loops with fixed, and Gaussian-distributed steplengths, in order
to look at the even broader distributions which allow for very long
segments, we also generated loops with the generalized
Cauchy-Lorentz ``random-flight'' distribution.  Finally, we
include loops of bimodally distributed fixed steplength.

In brief, our results are as follows.  First, we confirm the
exponential decay law of the unknot probability, formula
(\ref{eq:triv_P}), across all models examined.  Second, we find
qualitatively that indeed a wider distribution of the segment
lengths leads to knot suppression, ie a larger $N_0$.
Third, and most unexpectedly, we find that $N_0$ does not show any
signs of any singular behavior associated with the divergence of
mean squared segment length or any other moment of the segment
length distribution.  Instead, $N_0$ blows up and appears to grow
without a bound when the distribution of segment lengths
approaches the border of normalizability.

\begin{figure}
\centerline{\scalebox{0.4}{
\includegraphics{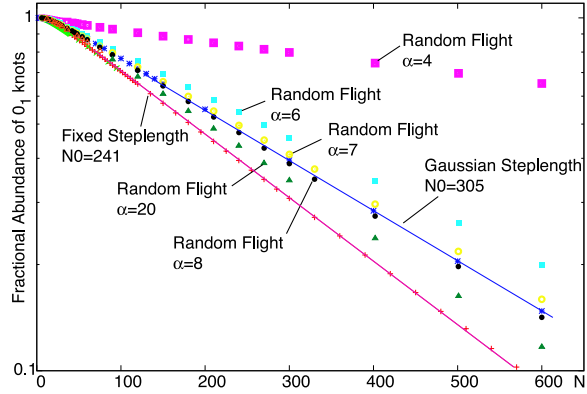}
}}\caption{(Color online) The fraction of loops of $N$ segments
with trivial topology ($0_1$, or unknots) follows the decaying
exponential form given in eq.(\protect\ref{eq:triv_P}), as seen in
this semi-logarithmic plot. The value of the decay constant $N_0$
varies considerably as the loop structure is changed. This figure
displays the fraction trivially knotted for loops of fixed or
gaussian-distributed steplengths. The incongruity of $N_0$ between
these two models, which for a variety of metrics are
indistinguishable at large loop lengths, is apparent. The figure
also shows data for a modified random-flight model, eq.
(\protect\ref{eq:mcl_prob}).  In this model, the value of $N_0$ is
tunable by way of the parameter $\alpha$, which allows a
substantial range of $N_0$ values, as indicated in the plot. The
other model studied has bimodal distribution of steplengths, as
described in section (\ref{section:bimodal_loops}), and although
not shown in the figure, allows similar variability in the decay
length $N_0$. } \label{fig:compare_fixed_and_gaussian}
\end{figure}

\section{Models and Simulation Methods}\label{sec:Methods}

\subsection{Models}

All polymer models referenced and employed in this work use a
freely jointed model to represent a polymer loop. The polymer is
represented by a set of $N$ vertices in 3D, with position vectors
$\vec{x}_i$, where the step between successive vertices is
described, $\vec{r}=\vec{x}_{i+1}-\vec{x}_{i}$.  In all models, we
assume that the distribution of segment vectors ${\vec r}$, which
we call $P({\vec r})$, is spherically symmetric and depends only on
the steplength $r = \left| {\vec r} \right|$, such that
\begin{equation} 
\left< {\vec r} \right> = \int {\vec r} P(r) d^3 r = 0 \ . 
\end{equation}
We also assume that the mean squared steplength is always the
same (when defined! - see below), we denote it $\ell$:
\begin{equation} \left< r^2 \right> = \int r^2 P(r) d^3
r = 4 \pi \int_0^{\infty} r^4 P(r) dr = \ell^2 \ . \end{equation}
With this in mind, the simplest measure of the distribution
breadth involves higher order moments:
\begin{equation}
\sigma^2 = \frac{\left< r^4 \right> - \left< r^2 \right>^2}{\left<
r^2 \right>^2} \  ,
    \label{eq:sigm_def}
\end{equation}
where $\left< r^4 \right> = 4 \pi \int_0^{\infty} r^6 P(r) dr$.

Specifically, we analyzed the following models.

The \textbf{fixed steplength model} is described by the distribution
\begin{equation}
P(\vec{r}) = \frac{\delta(|\vec{r}|-\ell)}{4 \pi \ell^2} \ .
\end{equation}
For this model, of course $\sigma=0$.  With $N$ segments, the loop's 
contour length is obviously $L=N\ell$ and the mean squared
gyration radius of the loop is $\left< R_g^2 \right>=(N+1)
\ell^2/12$.

The \textbf{gaussian steplength model} is generated by the
distribution
\begin{equation}
P(\vec{r}) = \left(\frac{3}{2\pi \ell^2}\right)^{3/2} \exp\left[-
\frac{3 r^2}{2 \ell^2}\right] \ ,
\end{equation}
in this case, $\sigma=\sqrt{2/3}$.  The contour length of the
$N$-segment loop in this model is $L = N \left< \left| {\vec r}
\right| \right> = N \ell \sqrt{8/3 \pi}$ and the mean squared
gyration radius is $\left< R_g^2\right> = \left( N-1/N
\right)\ell^{2}/12$.

The \textbf{random flight steplength model} is obtained from the
generalized Cauchy-Lorenz distribution (also known in the theory
of L\'evy flights \cite{Levy_flight}) of the form
\begin{equation}
P(\vec{r}) =  \frac{\left. \alpha \sin\left[3\pi/\alpha\right]
\right/ \left( 4 \ell^3 \pi^2 \right) }{1+(r/\ell)^{\alpha}} \ ,
\label{eq:mcl_prob}
\end{equation}
where the factor in the numerator ensures normalization.  Here,
$\alpha$, a parameter of the distribution, must be greater than
$\alpha > 3$, otherwise the normalization integral diverges.
Nevertheless, it would be fair to speak about a family of
random-flight models, parameterized by $\alpha>3$, instead of just
one model.  Varying $\alpha$ allows us to work with a ``tunable''
distribution.   These distributions' ``fat'' power-law tails lead
to diverging moments (which is why they are used to describe
super-diffusive behavior seen in biological foraging
\cite{Brazil_search,protozoa,fruit_flies,monkeys,deer,albatross},
and quite recently, in the diffusion of bank-notes across the
United States, \cite{Nature_money}).  Specifically, the
ensemble-averaged contour length of the loop is well defined only
at $\alpha > 4$ ($L =\left.  N \ell \sin \left(3 \pi / \alpha
\right) \right/ \sin \left( 4 \pi \alpha \right)$), mean squared
gyration radius exists at $\alpha > 5$, and $\sigma$ only exists
at $\alpha >7$, in which case it is equal to
\begin{equation}
\sigma= \sqrt{\frac { \sin\left[ 5 \pi / \alpha\right]^2 }{
\sin\left[ 3 \pi / \alpha \right] \sin\left[ 7\pi / \alpha
\right]}-1 } \ \ \ ({\rm at} \ \alpha > 7) \ .
\end{equation}
%
%Perceiving an incongruence between the fixed-step and
%gaussian-step loops, we also created loops with steplengths
%assigned from a distribution similar to the random-flight model of
%Cauchy and Lorentz, or in a broader sense, L\'evy.

Finally, we also include loops with \textbf{bimodally distributed
steplength}.  For these loops,
\begin{equation}
P(\vec{r}) = P_1\frac{\delta(|\vec{r}|-\ell_1)}{4 \pi \ell_1^2} +
P_2\frac{\delta(|\vec{r}|-\ell_2)}{4 \pi \ell_2^2} \ .
\end{equation}
which means that two possible steplengths, $\ell_1$ or $\ell_2$,
occur with probabilities $P_1$ and $P_2$, subject to the
normalization conditions, $P_1 + P_2 = 1$, and $\left< r^2 \right>
= P_1 \ell_1^2 + P_2 \ell_2^2 =\ell^2$.  All bimodally distributed
models can be conveniently parameterized by $P_1$ and $\lambda =
\ell_2/\ell_1$.  For these models,
\begin{equation}
\sigma  = \sqrt{\frac{P_1 + (1-P_1) \lambda^4}{(P_1 + (1-P_1)
\lambda^2)^2} - 1},
\end{equation}
might be very large if $\lambda$ is very large and $P_1$ is rather
close to unity ($\lambda^2 \gg 1/(1-P_1) \gg 1$).

\subsection{Loop generation}\label{sec:generation}

Unbiased generation of closed loops is of decisive importance for
our work.  Recently, we gave a detailed review of the existing
computational methods to generate statistically representative
closed loops (see the last section of the work \cite{Nathan_PRE}).  In
principle, the best way to generate loops is based on the
so-called conditional probability method.  The idea is that a
closed path is generated as a random walk, step by step, except
after the completion of $k$ steps, the next step, $k+1$, is generated from
the analytically computed probability distribution of the step
vector ${\vec r}$, subject to the condition that after $N-k$ more
steps, the walker returns to the starting point.  This idea was
first suggested and implemented for Gaussian distributed
steps \cite{vologodskii_gaussian_steps}. Recently, we
implemented this method for steps of equal length
\cite{Nathan_PRE}.  Unfortunately, this method is computationally
costly, and appears to be prohibitively difficult to implement for
more sophisticated models, such as random flight.

We therefore use the simpler method, called the method of triangles.
This method \cite{Nathan_PNAS,Nathan_Chapter_19,Nathan_PRE}
generates loops of $N$ segments with $N$ divisible by $3$. It
involves creating a set of $N/3$ equilateral triangles, each randomly
oriented in 3D space. Each triangle is considered a triplet
of vectors with zero sum.  A random permutation of the $N$ edge
vectors which make up these $N/3$ triangles, and then connecting
all $N$ vectors head-to-tail, creates a loop which will be 
closed, as the $N$ bond vectors together have $0$ vector sum.  Of course,
this method imposes correlations between segments.  We therefore
take special care to compare the results of this method with the
unbiased generation using the conditional probability method for
both Gaussian distributed and equal length step models.  We found
that no appreciable deviations in knot abundance data arise from
the imperfection of the triangle method.  We therefore use the
method of triangles to generate the random flight and bimodal
distributed loops, for which no alternative method is available.
To avoid even the slightest problems with correlations, implicit in
our triplet method, and to ensure that the decay of trivial knot
probability is in the exponential regime, we exclude the data from
small loops and fit the trivial knot probability on the interval
$N \in [50,300]$.

\subsection{Identification of topology and statistics}

For each of the models listed above, we generated loops of up to
at least $N=300$ segments.

Once the loop was generated, its knotted state was assessed
computationally with the Alexander determinant, $|\Delta(-1)|$, as
well as the Vassilev invariants of degree $2$ and $3$, $v2$ and
$v3$.  For this purpose, we employed knot analyzing routines
described in detail elsewhere \cite{Lua_Invariants}.

The fraction of generated loops with trivial topology was recorded
for each loop type.  In the interval $50 \leq N \leq
300$, every sample consisted of at least $10^6$ loops.  As a
result, the plot of trivial knot probability was created for each
model, with statistical error bars smaller than the data points in
Figure \ref{fig:compare_fixed_and_gaussian}.  Based on the data,
$N_0$ was measured for every model in our repertoire.

\section{Results}\label{sec:Results}

Our main results are shown in figure
\ref{fig:compare_fixed_and_gaussian}.  There, we present the
semi-log plots of the data on the trivial knot probability as a
function of the number of segments $N$ for a variety of models. To
begin with, all our data agree quite well with the exponential
character of the $w_{\rm triv} (N)$ dependence, formula
(\ref{eq:triv_P}).  Our main emphasis is therefore the study of
the characteristic value of $N_0$ for various models.

\subsection{Loops of fixed steplength versus Gaussian distributed steplength}
\label{fixed_step_loops}

Let us start with the two most commonly used models.  In the case
of fixed length steps, we obtain $N_0 \approx 240$ (see also
\cite{Nathan_PNAS}).  By contrast, for Gaussian distributed length
steps we get $N_0 \approx 305$ (see \cite{Nathan_PRE}).

\begin{figure}
\centerline{\scalebox{0.4}{
    \includegraphics{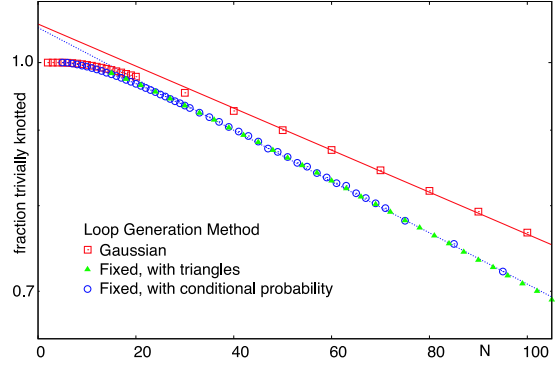} }}
\caption{
    (Color Online) Semi-log plots of trivial knot probability against the
    number of segments for the Gaussian distributed step length model
    ($\Box$), and for the fixed steplength model, the latter generated
    either via conditional probability method ($\circ$) or triangles
    method ($\triangle$).  The results indicate no dependence on the
    generation method, thus lending credence to the simpler method of
    triangles. The results also indicate an almost perfect fit to the
    exponential law, Eq. (\protect{\ref{eq:triv_P}}), with $N_0 \approx
    240$ for fixed steplength and $N_0 \approx 305$ for Gaussian distributed steps.
} \label{fig:checking_triangles}
\end{figure}

All data for the Gaussian model was generated using the fundamentally
unbiased conditional probability method.  For the model with fixed length
steps, we compare in figure \ref{fig:checking_triangles}
the data obtained by the conditional probability method
\cite{Nathan_PRE} and data generated by the much more efficient method
of triangles (see above section \ref{sec:generation}).  As the
figure indicates, there is practically no visible difference in
the results.  Accordingly, we unreservedly rely on the triangles
method in the rest of this work.

\begin{figure}
\centerline{\scalebox{0.4}{
    \includegraphics{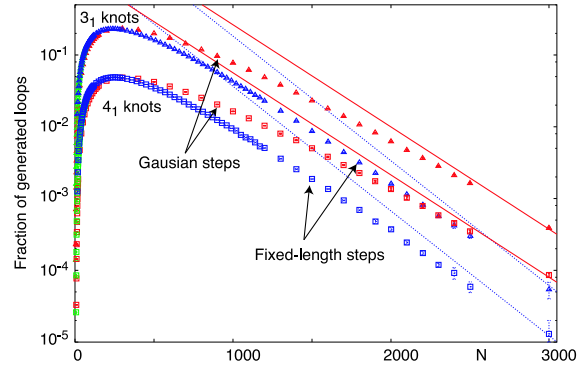}
}}
\caption{
    The probabilities of simple non-trivial knots $3_1$
    and $4_1$ are plotted against the number of segments in the loop in
    semi-log scale.  The data at large $N$ is consistent with
    the exponential form of these probabilities (plotted as solid lines in the figure), 
    which are characterized by the same parameter $N_0$ found for trivial knots.  
    As with the case of trivial knots, we see the difference between the fixed steplength
    model and Gaussian distributed model, namely, that any particular knot
    probability decays slower for Gaussian distributed segments than for fixed segments.
}
\label{fig:other_knots}
\end{figure}

Although our main attention in this paper is on the trivial knots
(unknots), we show in figure \ref{fig:other_knots} some data for
more complex knot probabilities.  Our data at least do not
contradict the assertion that the probability of every particular
knot decays exponentially at sufficiently large $N$, with the same
decay length as the trivial knot probability.  At the same time,
our data also confirm the systematic difference between models -
decay length, although the same for all types of knots, does
depend on the distribution of step lengths involved.  Specifically,
probabilities of knots $3_1$ and $4_1$ decay at a noticeably slower
rate for the Gaussian model than for the model with fixed steps.

\begin{figure}
\centerline{\scalebox{0.4}{ \includegraphics{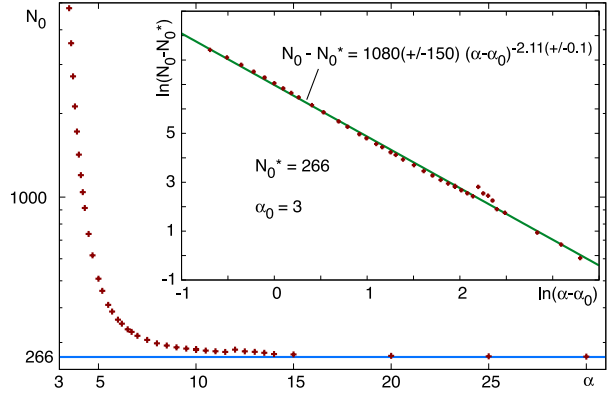} }}
\caption{
    (Color Online) Values of the trivial knot
    decay length, $N_0$, measured from loops constructed of steps
    with generalized Cauchy-Lorenz distribution (reminiscent of
    L\'evy-flights).  The decay constant is expressed as dependent on
    the variable $\alpha$ which defines how fat the tail of the
    distribution remains at large steplengths.  The knot probability
    decay constant, $N_0$, approaches a constant value of about $266$ at large
    $\alpha$.  The inset shows the same data in double logarithmic
    scale. It is seen that $N_0$ blows up approximately as
    $(\alpha - 3)^{-2}$ when $\alpha$ decreases. It is interesting to
    note that the dependence of $N_0$ on $\alpha$ does not show any
    signs of irregularity as $\alpha$ crosses values at which various
    moments of segments length distribution start diverging (for
    instance, mean squared gyration radius diverges at $\alpha \leq 5$
    and mean contour length diverges at $\alpha \leq 4$).
}
\label{fig:mcl_fit}
\end{figure}

\subsection{Random flight loops}

The results of our study of loops with random-flight steplength,
$\alpha \in [3.5,30]$, are summarized in figures
\ref{fig:compare_fixed_and_gaussian} and \ref{fig:mcl_fit}. The
raw curves of probability given in figure
\ref{fig:compare_fixed_and_gaussian} clearly show that the odds of
finding an unknot in a set of loops get increasingly unfavorable
as $\alpha$ decreases.  That is not unexpected: at smaller
$\alpha$, the probability distribution (\ref{eq:mcl_prob}) acquires
an increasingly fat tail, which implies the presence of a fraction of
exceptionally long segments, and they of course suppress the
chance of knots.

At very large $\alpha$, the knot probability for the random flight
model appears similar to the data for fixed steplength loops.  
Indeed, as figure \ref{fig:mcl_fit} indicates,
$N_0$ for the random flight model at very large $\alpha$
approaches $N_0 \approx 266$ which is not dramatically
different from $N_0 \approx 240$ for the fixed length steps.  In
fact, the remaining difference might be associated with the fact
that even at very large $\alpha$, the random flight model, although it
has essentially no very long steps, has some relatively short
ones, which might account for the discrepancy in $N_0$.   
Figures \ref{fig:compare_fixed_and_gaussian} and
\ref{fig:mcl_fit} show further that at $\alpha \approx 7.5$, the
random-flight loops behave essentially the same way as Gaussian steps in
terms of $N_0$.

Most interestingly, figure \ref{fig:mcl_fit} shows no sign of
anything unusual happening to $N_0$ at the values of $\alpha$ at
which various physically important moments of the segment length
distribution (\ref{eq:mcl_prob}) start diverging.  For instance,
at $\alpha \leq 5$ the mean squared gyration radius diverges, at
$\alpha \leq 4$ even the contour length of the loop diverges - and
yet none of these facts find any visible reflection on the
dependence of $N_0$ on $\alpha$.  $N_0$ keeps smoothly increasing
with $\alpha$, with a maximum measured value of 
$N_0 \approx 4800$ at $\alpha=3.5$.  It appears that $N_0$ in fact
blows up and goes to infinity as $ \alpha$ approaches $3$ - the
border below which the distribution (\ref{eq:mcl_prob}) is not
normalizeable.  Moreover, as the inset of figure \ref{fig:mcl_fit}
shows, this divergence is well approximated by the power law
dependence of the form,
\begin{eqnarray} N_0 = N_0^{\ast} &+& B
(\alpha-\alpha^{\ast})^{- \beta} \ , \ \ {\rm where} \nonumber
\\  N_0^{\ast} & \approx & 266 \ , \ B \approx 1080 \pm 150 \ , \alpha^{\ast} =3 \ , \
\ {\rm and} \nonumber \\  \beta & \approx & 2.1 \pm 0.1 \ .
\end{eqnarray}
In fitting the data with this power law, we ignored the small
irregularities visible around $\alpha\approx13$ (or $\ln \left[
\alpha-\alpha_0 \right] \approx 2.2$), which we attribute to the
numerical problems with the cutoff implemented in our use of
eq.~(\ref{eq:int_prob}).

The meaning and physical origin of the apparent criticality
observed at $\alpha$ approaching its minimal possible value of $3$
currently evades our understanding.

\subsection{Loops of bimodal-distributed steplength}
\label{section:bimodal_loops}

For this model, we examined the interval $N\in[50,600]$, with at
least $10^6$ loops in each simulation record for $N\le300$, and at
least $10^5$ loops in those records used with length $N>300$.  We
also consider parameters $P_1$ and $\lambda = \ell_1/\ell_2$ in
the intervals $P_1 \in [0,1]$ and $\lambda \in [0.01,1.0]$,
respectively. By symmetry, it is sufficient to look at $\lambda <
1$; this means, $P_1$ is the fraction of \emph{shorter} segments.

\begin{figure}
\centerline{ \scalebox{0.4}{  \includegraphics{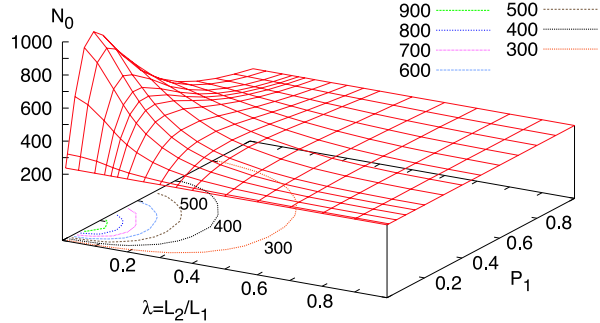} }}
\caption{
    (Color Online) The surface plot of $N_0$ measured for
    loops of bimodally distributed steplength.  The two steplengths
    used, $\ell_1$ and $\ell_2$, related by $\lambda = \ell_2/\ell_1$,
    occur with probabilities $P_1$ and $P_2$.  The surface is
    described by two degrees of freedom, $\lambda$ and $P_1$.  The
    model reduces to the fixed steplength model of section
    \ref{fixed_step_loops} in the conditions $P_1=0$, or $P_1=1$, or
    $\lambda=1$, as is seen by the reduction of $N_0 \to 241$ in these
    limits.  Interestingly, the bimodal distribution of steplengths
    never reduces $N_0$ below the limit of $N_0=241$ of the
    fixed-steplength model.  The maximum value of $N_0$ which we were
    able to observe, $N_0=974$, occurs at $P_1=0.15$, $\lambda=0.01$.
}
\label{fig:bimodal_surface}
\end{figure}

The raw data from the loops with bimodal steplength are presented
in figure \ref{fig:bimodal_surface}. This surface plot charts the
change in $N_0$ as a function of two parameters of the model,
$P_1$ and $\lambda$.  Below the surface is the corresponding
contour plot of $N_0$ as a function of these two parameters. The
changes in $N_0$ are smooth, and there do not seem to be any
singularities in the behavior of $N_0$.  The maximum observed
value of $N_0=974$ occurs at $P_1=0.15$, $\lambda=0.01$. This
maximum of $N_0$ appears to be rather sharp, small deviations in
$P_1$ from $0.15$ lead to smaller $N_0$ values. From the data, we
believe that in this model $N_0$ is maximized when the fraction of
shorter segments is $P_1=0.15 \pm 0.05$. As regards the second
degree of freedom, $\lambda$, while our data certainly shows that
as $\lambda\to1$ the bimodal system approaches the fixed
steplength system of section \ref{fixed_step_loops} (ie
$N_0=241$), in the opposite direction of $\lambda\to0$, it is not
clear if $N_0$ will continue to increase in an unbounded way or be
in some way encumbered. In this regard our present knot analysis
machinery is limited by the relative disparity between segments of
different length, and more work needs to be done to elucidate the
scaling of $N_0$ in the limit $\lambda \to 0$.

\begin{figure}
\centerline{ \scalebox{0.4}{ \includegraphics{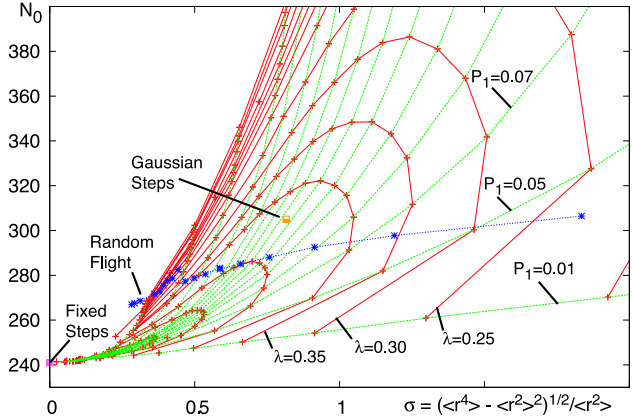}}}
\caption{
    (Color Online) The value of $N_0$ for several different loop
    models is displayed against the steplength distribution width
    $\sigma$, which is defined in eq. (\protect\ref{eq:sigm_def}). The
    data in this figure fail the hypothesis that a larger $\sigma$
    universally maps to a larger $N_0$. The measurement for fixed
    steplengths, shown in the lower left corner is the point from
    which all bimodal loop data emanates.  As labeled in the figure,
    the manifold for bimodal steplength is indexed by lines of
    constant $P_1$ and $\lambda$.
}
\label{fig:bimodal_compare_small}
\end{figure}

Qualitatively, all of our data are consistent with the idea that what
suppresses knots is the presence of a fraction of unusually long
segments.  One could then hypothesize that $N_0$ might depend on
some unique property of the segment length distribution,
for instance, $\sigma$, as defined in eq. (\ref{eq:sigm_def}).  This hypothesis is
tested in figure \ref{fig:bimodal_compare_small}; the figure
indicates that the hypothesis fails.  Nevertheless, the results
presented in this figure are interesting, as they show that large
$N_0$ can be achieved by the combination of segment length
difference ($\lambda$) and segment type fractions.

Although a functional relationship between $N_0$ and $\alpha$
seems evident in our data for random flight loops, the data we
have for loops with bimodal distribution of steplength do not
suggest a simple single parameter which determines $N_0$.

% proofed to here

\section{Conclusions}\label{sec:Conclusion}

In a qualitative sense, it does seem that there exists a relationship
between $N_0$ and the reach of successive segments within the
chain (as seen in Figure \ref{fig:bimodal_compare_small}).  It
seems qualitatively clear indeed that knottedness is greatly
suppressed by the presence of some very long segments. Thus, the
slowly decaying tail of the segment length probability distribution,
or the presence of a small fraction of very long segments, implies
a large $N_0$, which sounds natural. However, our understanding of
these observations beyond the qualitative level is limited.  We
found that $N_0$ appears to exhibit no singularity associated with
divergence of any natural characteristics of the loop, such as its
gyration radius or even contour length; instead, $N_0$ exhibits
power law critical behavior when the segment length distribution
approaches the boundary of normalizability.  In addition, we do
not know which property of the segment length distribution
determines $N_0$.  We have established that this is not simply the
distribution width, and from the results with random flight
distribution it seems also clear that it is not based on any
finite moments of the distribution. We consider the development of
a fuller understanding of the variance of $N_0$ a compelling
challenge.

The data clearly show that wide variation in the behavior of $N_0$
is possible, including very large values of $N_0$ in certain
models.  Given that $N_0$ plays the role of the cross-over length
for the critical behavior of topologically constrained loops
\cite{Nathan_PRE}, we can speculate that the models with very large
$N_0$ are in some way similar to the models of self-avoidance in
the vicinity of the $\theta$-point.  We think that this analogy
deserves very close attention.

\subsection*{Acknowledgments}

We acknowledge useful discussion with A.L.~Efros and A.~Vologodskii.
We express thanks to R. Lua for the use of his Knot Analysis routines.
We are grateful for access to the
Minnesota Supercomputing Institute's IBM Core Resources and the
University of Minnesota Physics Department linux cluster,
resources which made much of the simulation possible.  This work
was supported in part by the MRSEC Program of the National Science
Foundation under Award Number DMR-0212302.

\appendix

\section{On the numerical implementation of generalized
Cauchy-Lorenz distribution}

To generate steplengths from the random-flight distribution
(\ref{eq:mcl_prob}), we first take a random number $q$ from the
uniform distribution on the interval $[0,1]$ and then find
steplength $r$ as $r=f(q)$, where the mapping $f(q)$ is determined by the
equation
\begin{equation}
q = \int_0^{f(q)} P(\vec{r}) 4 \pi r^2 d r \ , \label{eq:int_prob}
\end{equation}
where $P(\vec{r})$ is given by eq. (\ref{eq:mcl_prob}). Although a
closed-form representation of the right-hand side of eq.
(\ref{eq:int_prob}) exists in the form of an Incomplete Beta
Function, \cite{incomplete_beta}, we chose to implement the
mapping via a numerical interpolation of tabulated values of the
integral. As the power-law tail of this distribution is quite fat,
particularly at small $\alpha$, accurate representation of the
integral becomes challenging at large steplengths.  This work is
ultimately numerical, and we are forced to truncate the
representation of the integral to $1.0$ at some maximum
steplength, ie specifying an upper bound on $f(q)$. The slow
convergence of the tail is behind the appearance of the few
outliers in the data for the random-flight method, figure
\ref{fig:mcl_fit}.

\end{document}